\documentclass[]{spie}  

\usepackage{amsmath,amsfonts,amssymb}
\usepackage{graphicx}
\usepackage[colorlinks=true, allcolors=blue]{hyperref}
\usepackage{float}

\title{The Versatile CubeSat Telescope: Going to Large Apertures in Small Spacecraft}

\author[a]{Jaren N. Ashcraft}
\author[b]{Ewan S. Douglas}
\author[a,b,c]{Daewook Kim}
\author[a]{George A. Smith}
\author[d]{Kerri Cahoy}
\author[b]{Tom Connors}
\author[a]{Kevin Z. Derby}
\author[b]{Victor Gasho}
\author[b]{Kerry Gonzales}
\author[a]{Charlotte E. Guthery}
\author[e]{Geon Hee Kim}
\author[b]{Corwyn Sauve}
\author[d]{Paul Serra}

\affil[a]{James C. Wyant College of Optical Sciences, University of Arizona, Meinel Building 1630 E. University Blvd., Tucson, AZ. 85721}
\affil[b]{Department of Astronomy and Steward Observatory, University of Arizona, 933 N. Cherry Ave., Tucson, AZ 85719, USA}
\affil[c]{Large Binocular Telescope Observatory, University Of Arizona, 933 N. Cherry Ave. Tucson, AZ 85721}
\affil[d]{STAR Lab, Department of Aeronautics and Astronautics, Massachusetts Institute of Technology, Cambridge, MA 02139}
\affil[e]{Department of Mechanical Material Convergence System Engineering, Hanbat National University, Daejeon, South Korea}
\authorinfo{Further author information: (Send correspondence to J.N.A.)\\J.N.A.: E-mail: jashcraft@email.arizona.edu\\  E.S.D.: E-mail: douglase@email.arizona.edu}

\begin{document} 
\maketitle

\begin{abstract}
The design of a CubeSat telescope for academic research purposes must balance complicated optical and structural designs with cost to maximize performance in extreme environments. Increasing the CubeSat size (eg. 6U to 12U) will increase the potential optical performance, but the cost will increase in kind. Recent developments in diamond-turning have increased the accessibility of aspheric aluminum mirrors, enabling a cost-effective regime of well-corrected nanosatellite telescopes. We present an all-aluminum versatile CubeSat telescope (VCT) platform that optimizes performance, cost, and schedule at a relatively large 95 mm aperture and 0.4 degree diffraction limited full field of view stablized by MEMS fine-steering modules. This study features a new design tool that permits easy characterization of performance degradation as a function of spacecraft thermal and structural disturbances. We will present details including the trade between on- and off-axis implementations of the VCT, thermal stability requirements and finite-element analysis, and launch survival considerations. The VCT is suitable for a range of CubeSat borne applications, which provides an affordable platform for astronomy, Earth-imaging, and optical communications. 
\end{abstract}


\section{INTRODUCTION}

CubeSat space telescopes are a growing asset in astronomical research\cite{Shkolnik_2018,douglas2019cubesats}. Their standardized format and compact volume enables the rapid development of high-impact mission concepts above the earth's atmosphere. Flight-tested commercial microelectromechanical systems (MEMS)  have further enhanced CubeSat telescopes by providing dynamic pointing and aberration control on-orbit (DeMi, NODE) \cite{Yenchesky2019OptomechanicalDA,Clements2016}. The expanding capacity of CubeSats for astronomical research invites the development of novel technologies for their optical systems. However, the resolution in CubeSat optics is fundamentally limited by the clear aperture permitted by the CubeSat volume. The rayleigh criterion tells us that resolution ($\Delta \theta$) is inversely related to to the entrance pupil diameter ($D$) of the optical system.

\begin{equation}
    \Delta \theta = 1.22 \frac{\lambda}{D}
\end{equation}

Reflective CubeSat objectives traditionally require mounting hardware that limits the available entrance pupil diameter. Single-point diamond turned (SPDT) mirrors can mitigate this limitation by enabling the manufacturer to machine the mounting hardware directly into the rear of the mirror substrate, eliminating the need for mounting hardware around the edge of the mirror and enhancing the nominal throughput and resolution. This also grants the objective a considerable degree of athermalization by eliminating the difference in the coefficient of thermal expansion between the mirror and mounting hardware\cite{Zhang:17}. SPDT surfaces are typically used in longer wavelengths (MWIR, LWIR) due to the midspatial frequency errors left by tooling marks. However, recent developments in optical polishing\cite{Jeon:17} have introduced Magnetorheological Finishing (MRF) to the fabrication of SPDT surfaces, reducing the dominant midspatial frequencies considerably and therefore extending their use into the optical. Athermal CubeSat objectives composed of low surface roughness mirrors with a large entrance pupil invites the design of high-performance optical CubeSat payloads to make the next generation of research and technology development in space more accessible. The Versatile CubeSat Telescope (VCT) concept is a prototypical fore-optic that boasts compatibility with a variety of research payloads. The goal of the VCT is to develop a flexible, large aperture telescope that can be replicated at low cost for easy adaptability to future research payloads and technology demonstrations. 

\section{OPTICAL DESIGN - ON-AXIS V.S. OFF-AXIS IMPLEMENTATIONS}
\begin{figure}[H]
    \centering
    \includegraphics[width=0.9\textwidth]{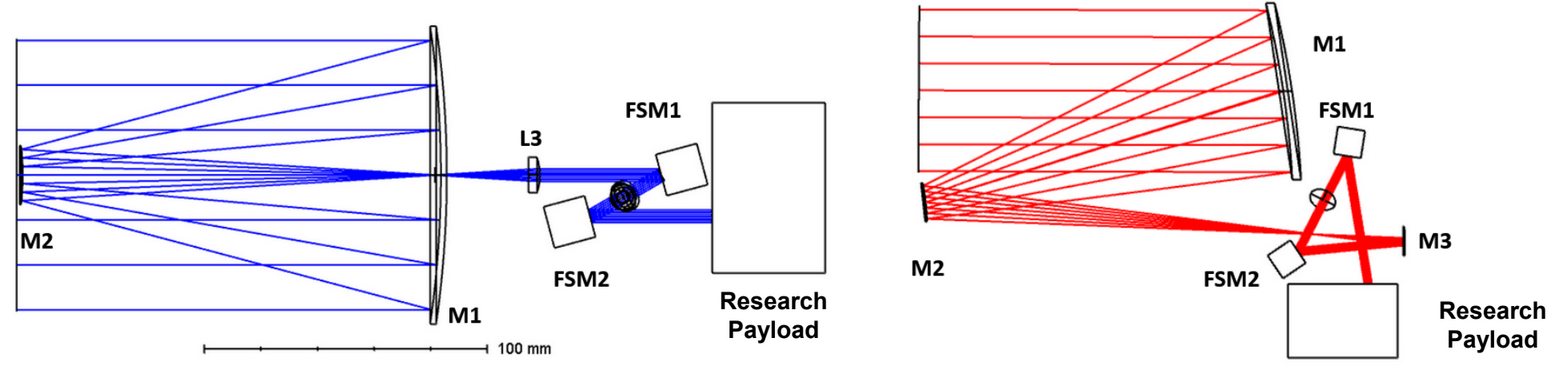}
    \caption{System layouts of the (left) On-Axis Telescope and the (right) Off-Axis Telescope designs. Each has three optical elements, two fast-steering mirrors (FSM) for fine pointing control, and a volume dedicated to a research payload.}
    \label{fig:layout}
\end{figure}

The design of the VCT began with two realizations of a CubeSat telescope that could take advantage of a large primary mirror. The first, an on-axis telescope with a more classical Ritchey-Chretien objective outfitted with a plano-convex aspheric collimator. The second, an all-reflective off-axis Ritchey-Chretien solution with a freeform collimator. We baselined a 95mm entrance pupil diameter with a 20$\%$ obscuration for the on-axis telescope, and scaled the entrance pupil diameter of the off-axis telescope to be equivalent in collecting area. A high pupil magnification is required for both designs to image the primary mirror onto the small MEMS FSM clear aperture (5mm). The specifications and system layouts are shown in table \ref{tab:specs} and figure \ref{fig:layout} respectively.

\begin{table}[H]
    \centering
    \begin{tabular}{c c}
        \hline
        Specification & Value  \\
        \hline
        Entrance Pupil Diameter & \textbf{95 mm} \\
        Exit Pupil Diameter & 5 mm \\
        Half Field of View & 0.2$^{\circ}$ \\ 
        Secondary Obscuration & \textbf{20$\%$} \\
        Central Field Strehl & $>$0.99 \\
        Average Field Strehl & $>$ 0.80 \\
        OTA Packaging Volume & $<$ 2U \\
        \hline
        \\
    \end{tabular}
    \qquad 
    \begin{tabular}{c c}
        \hline
        Specification & Value  \\
        \hline
        Entrance Pupil Diameter & \textbf{88 mm} \\
        Exit Pupil Diameter & 5 mm \\
        Half Fiel of View & 0.2$^{\circ}$ \\ 
        Secondary Obscuration & \textbf{0$\%$} \\
        Central Field Strehl & $>$0.99 \\
        Average Field Strehl & $>$ 0.80 \\
        OTA Packaging Volume & $<$ 2U \\
        \hline
        \\
    \end{tabular}
    \caption{Optical design specifications for the (left) On-Axis and (right) Off-Axis telescope designs. All specifications are similar except for those in bold, which are set such that the apertures of each design are equal in throughput.}
    \label{tab:specs}
\end{table}

 Aspheric surfaces were necessary for both telescopes to achieve a high pupil magnification (19x and 17.5x for the on- and off-axis designs respectively) while maintaining a well-corrected field of view. While this slightly complicates the metrology of the mirrors it did not add any additional cost because the mirrors were already going to be diamond-turned. Both telescopes utilized a Ritchey-Chretien objective that forms an image near the primary mirror, which is collimated by an aspheric element to produce an external exit pupil. The size of the exit pupil was chosen to match the clear aperture of a MEMS fast-steering mirror (FSM)\cite{memsfsm,Serra21} to correct for spacecraft pointing and jitter errors while in orbit. The external pupil also serves as a convenient interface for any instrument suite that would be included on the VCT (e.g. spectrographs, cameras, lasers).  The concept for the on-axis design was to make a well-established telescope format work given the pupil magnification and field of view specifications. The aspheric surfaces granted the design a well-corrected field of view in a compact format suitable for a CubeSat (2U). However, the design has a secondary obscuration which limits the throughput of the system and adds diffraction features to the image. The primary goal for the off-axis design was to offer a viable alternative to the classical on-axis design that would not suffer from undesirable diffraction effects brought upon by the secondary obscuration. Unfortuantely, shifting the primary parabolic mirror off-axis comes at the cost of more field-dependent aberration that limits the system’s field of view. To accommodate this we selected the tertiary mirror of the off-axis design to be freeform. Before optimization of the freeform surface, the rotational symmetry of the system was broken by tilting the secondary and tertiary mirrors about their foci to make the system free of linear astigmatism. A linear-astigmatism free three mirror system (LAF-TMS) has been shown by Park et al\cite{park_development_2020} to be an excellent starting point for freeform designs by using the mirror tilt angles and inter-mirror distances (Eq \ref{eq:laf}) to mitigate a dominating aberration early in the design process.

\begin{equation}
    \frac{l_2^\prime}{l_2}\frac{l_3^\prime}{l_3}tan{i_1}+\left(1+\frac{l_2^\prime}{l_2}\right)\frac{l_3^\prime}{l_3}tan{i_2+}\left(1+\frac{l_3^\prime}{l_3}\right)tan{i_3=0}
    \label{eq:laf}
\end{equation}

In this equation, parameters i$_{1,2,3}$ are the angles of incidence of the optical axis ray on surfaces 1,2, and 3, while $l_{2,3}$ and ${l^{\prime}}_{2,3}$ are the front and rear focal lengths the secondary and tertiary mirror. Once the baseline LAF-TMS design had been implemented, the third mirror was changed to a freeform XY-Polynomial surface where coefficients were optimized that maintained the bilateral symmetry of the optical system. The mirrors were shifted off-axis in the $\hat{y}$ direction, as was the field bias (+0.2$^{o}$). Consequently the freeform surface was constrained to be plane-symmetric about the $\hat{y}-\hat{z}$ plane by solely optimizing coefficients of even order in $\hat{x}$. The resultant design is nearly diffraction-limited across the biased field of view.

\begin{figure}[H]
    \centering
    \includegraphics[width=0.8\textwidth]{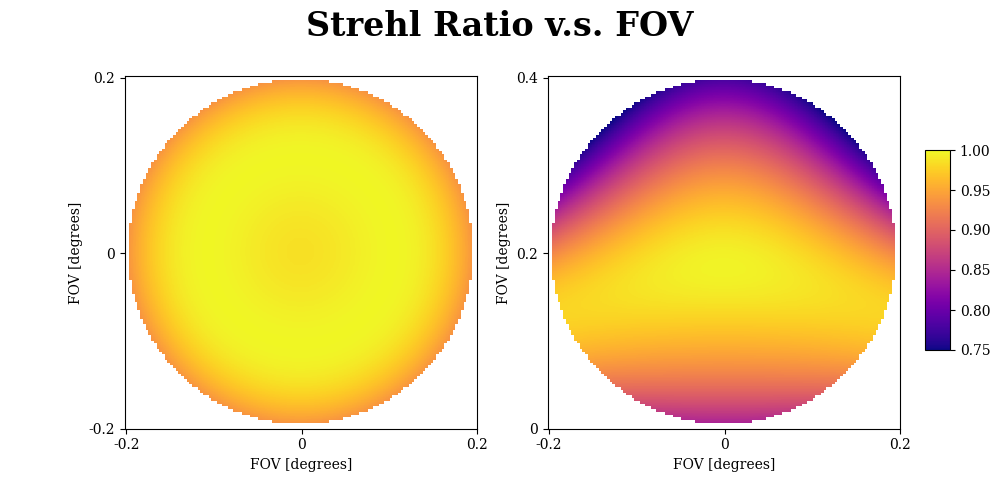}
    \caption{Maps of strehl ratio v.s. field of view for the (left) On-Axis and (right) Off-Axis VCT. The symmetry of the optical system mirrors the performance quite readily, with the on-axis design having a rotationally symmetric field performance whereas the off-axis design exhibits plane-symmetric performance.}
    \label{fig:strvfov}
\end{figure}

\section{SENSITIVITY ANALYSIS}
 
The misalignment sensitivities of the two designs were explored in Zemax OpticStudio (ZOS) by iteratively perturbing each optic in six degrees of freedom using the Python ZOS-API. Our as-built performance goal for both systems was an average Strehl ratio $>$ 0.7 across the field of view to maintain reasonably diffraction-limited performance. The $0\%$, $70\%$, and $100\%$ field of view in $\pm \hat{x}$ and $\pm \hat{y}$ were considered in the average calculation, considerably biasing the sensitivity analysis toward the edge of the field of view where performance is expected to degrade faster. This results in a more pessimistic analysis of misalignment sensitivity. The sensitivity analysis fixes the primary mirror while M2 and L3/M3 were perturbed to better understand how misalignments with respect to the primary would impact the optical performance over all fields of view.

\begin{figure}[H]
    \centering
    \includegraphics[width=\textwidth]{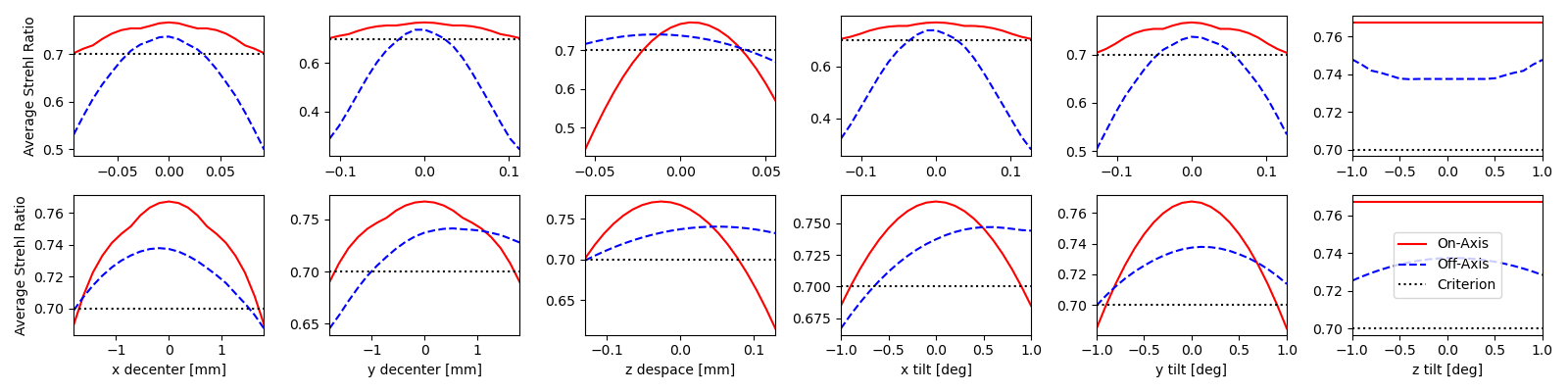}
    \caption{Sensitivity to misalignment curves of the On-Axis (Solid Red) and Off-Axis (Dashed Blue) designs for their respective secondary (top) and tertiary (bottom) optic. The on-axis system is less sensitive for X/Y decenter of both elements, X/Y/Z tilt of M2, and X/Z tilt of M3.}
    \label{fig:my_label}
\end{figure}

Figure 2 shows that in most cases the on-axis design was less sensitive to misalignment in six degrees of freedom. This result is consistent with the complexity of each system. The off-axis design employs off-axis conics and a freeform surface, both of which will be sensitive to misalignment due to the higher mirror slopes than the on-axis design. In applications where the secondary obscuration is tolerable, the on-axis design would be a better choice for applications where cost and misalignment risk are limiting factors.

\section{POLARIZATION ANALYSIS}
Many spaceborne optical payloads have a degree of sensitivity to polarization effects. From laser experiments in support of quantum communications\cite{Serra21} to polarimetry of atmospheric ice\cite{Hart20}, there is a clear need to characterize the response of cubesat payloads to a vector electric field. To offer support for polarization-sensitive research payloads a comprehensive polarization ray trace (PRT) analysis was conducted for each realization of the VCT.  PRT is a ray-based approach to vector field calculations enabled by determining the local fresnel reflection coefficients that alter the transmission and phase of the electric field at each surface in the optical system. These effects are traced from the entrance pupil to the exit pupil of the optical system to determine the cumulative polarization behavior of the instrument. The proposed VCT designs use fast primary mirrors to achieve the high pupil magnification, so some polarization aberration is expected. However mirror substrates and coatings can be selected to mitigate the impact of polarization aberration for a given application. 

\begin{figure}[H]
    \centering
    \includegraphics[width=\textwidth]{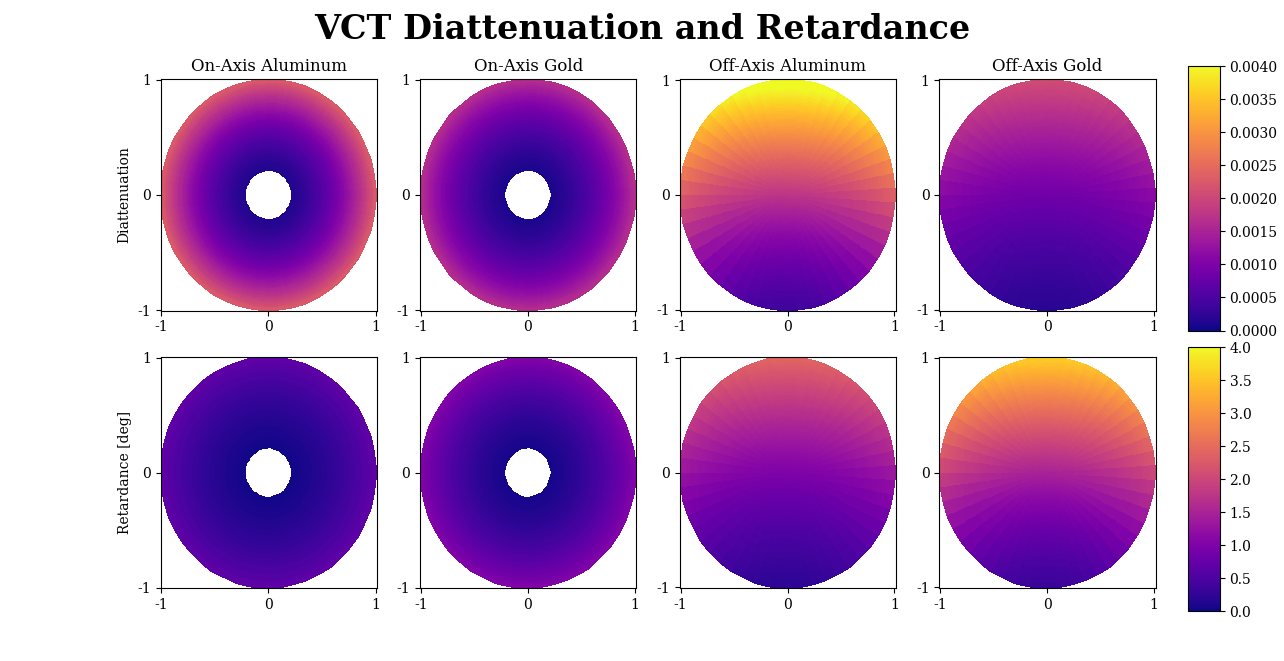}
    \caption{Polarization pupil maps assuming bare aluminum mirrors and uncoated N-BK7 lens surfaces. (Top) Diattenuation and (Bottom) retardance maps of the (left) On-Axis and (right) Off-axis versatile cubesat telescopes. These data were computed at $\lambda$ = 780nm. At this wavelength the Off-Axis Aluminum VCT sees the greatest diattenuation, and the Off-Axis Gold sees the greatest retardance. This indicates that the off-axis design would introduce more challenges to polarization-sensitive payloads than the on-axis design would.}
    \label{fig:polmaps_al}
\end{figure}

A PRT tool was written using the ZOS-API to create maps of diattenuation and retardance at the exit pupil of the system. The ZOS-API tool creates a batch raytrace and propagates it from the entrance pupil of the optical system to the surface under investigation. The output of the raytrace gives surface normal vectors in addition to exit ray vectors which are used to calculate the incident ray vectors and angle of incidence. Diattenuation ($D$) and retardance ($\delta$) data is produced from the Zemax coating files and ray angles of incidence to generate the surface and pupil maps shown in figure \ref{fig:polmaps_al}.

\begin{equation}
    D = \frac{|r_p|^{2} - |r_p|^{2}}{|r_s|^{2} + |r_p|^{2}};\phantom{flapjackfacts}\delta = |\phi_{p} - \phi_{s}|
    \label{eq:diat}
\end{equation}

The polarization pupil map permits analysis of the optimal material and configuration for a given polarization-dependent application. Figure \ref{fig:polmaps_al} demonstrates that aluminum is more diattenuating than gold at $\lambda=780nm$, and the off-axis configuration performs worse than the on-axis configuration due to the higher angles of incidence on the mirror surfaces. 

\section{MECHANICAL DESIGN}

The VCT must be designed to deliver a high optical wavefront when operational on-orbit, yet also robust enough to survive the rigors of launch and orbital injection separation.  Achieving both goals in a compact package that affords straight-forward assembly and alignment is quite challenging.  Another important aspect of the design must also consider the mass contribution of the structural elements as this can be costly when considering launch requirements and on-orbit maneuvering.  With the prevalence of small satellite design ( $<$ 12U), development of a telescope that fits within a 2U volume with a mass less than 2 kg would present a low-cost solution that enables rapid prototyping of research payloads. For a prototype realization of the VCT, the optical system design called for SPDT Aluminum primary and secondary mirrors. All mechanical structural elements were also chosen to be aluminum to match the thermal behavior of the mounting structure to the optical elements. The collimating lens, which locates the instrument pupil, could also be easily mounted to the OTA using a common aluminum barrel structure with simple shims for precise alignment. 

The first goal in the mechanical design of the telescope Optical Tube Assembly (OTA) is to support the optical elements with minimal effect due to expected thermal perturbations that would be encountered on-orbit.  These effects would include both rigid body misalignment and wavefront aberrations caused by subtle flexing of the optical elements.  In order to bound the problem, the first design decisions centered on overall sizing to fit within the small-sat form factor. The primary-secondary despace tolerance is very tight, so a truss structure was designed to properly space the two mirrors. The truss structure presents a good strength to weight ratio and is commonly used in ground-based telescope designs. For this space-based application, the truss structure would not allow for simple and easy shrouding of the optical path and thus would likely require additional structural pieces to support stray light baffling elements and thermal mitigation elements (i.e., mylar blankets).  Therefore, an early design decision was made to investigate a tube structure that would serve as both the metering structure that could maintain the mirror separation tolerance, transverse misalignment tolerance, and provide an in-place baffle that could be customized for the particular optical application.  A significant benefit of the tube design is the mass savings realized by the ability to select a very thin tube wall thickness that still retains the desired structural performance for all loading cases.  This tube would also allow for easy installation of thermal mitigation measures and targeted design of thermal conduction paths that further ensure optical performance stability while on-orbit.

The next major goal in the mechanical design considers the manufacturability of the system.  Considering its small size, the OTA manufacturability includes the fabrication processes for all structural components, as well as thoughtfully planned assembly and alignment processes that ensure the optical performance requirements can be met.  The aluminum mirrors provide a unique opportunity to meet the need for an assembly that is easy to integrate and align.  The primary mirror optical surface is thus applied directly to the structural component (primary mirror) which has appropriate design elements that consider thermal performance and mounting points for both the metering tube and the interface to the downstream optical system via a \emph{hex plate} interface.  Clearly defining the rear of the primary mirror blank as the OTA support location allows for simple design of flexural components that will not propagate errors from other systems into the OTA, or vice-versa.  Similarly, the secondary mirror surface is applied directly to a head-ring structure that also considers thermal performance and mounting points to the metering tube, but also minimizes the size of support strut spiders that are common in the on-axis  design.  Considering the small size of the OTA, the one-piece secondary mirror/head-ring simplifies its alignment features by moving them out to the tube diameter, rather than trying to squeeze them into the shadow of the secondary mirror on the entrance aperture.

\begin{figure}[H]
    \centering
    \includegraphics[width=0.55\textwidth]{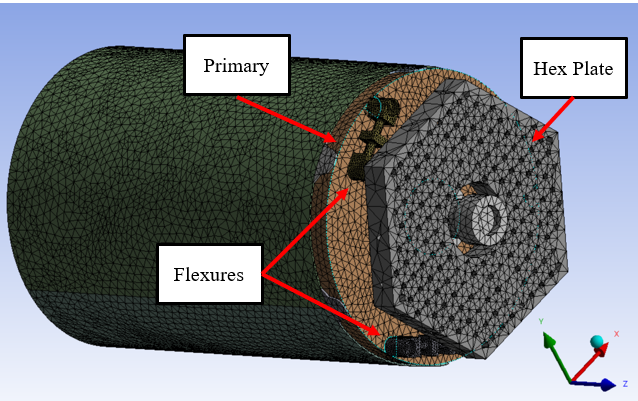}
    \caption{The FEM of the VCT OTA highlighting the hex plate, flexure, and tube design.}
    \label{fig:vct_fea}
\end{figure}

The final goal of the mechanical design considers the interface of the OTA to the instrument and the spacecraft.  Most terrestrial optical system applications utilize a breadboard style support that allows for many mounting possibilities for both the OTA and any downstream instrument opto-mechanical elements, and can be purposefully designed to incorporate spacecraft mounting options that will isolate the payload from the spacecraft both thermally and dynamically as necessary.  The hex plate provides all of these interfaces in a single machined aluminum component.  The connection of the OTA to the hex plate is of particular importance for this discussion.  As mentioned earlier, the OTA must remain dynamically and thermally isolated from the other spacecraft or instrument components to preserve the telescope optical performance. A Finite Element Model (FEM) was created using the aluminum tube structure OTA, with both structural and optical components represented, and a Finite Element Analysis (FEA) was performed to determine the model behavior with both structurally dynamic boundary conditions and also varying thermal environmental conditions.  The FEM is shown in figure \ref{fig:vct_fea}.

A simple three flexure design was selected to mount the OTA to the hex plate.  The flexures were positioned tangentially to the primary mirror blank to better accommodate thermal deformations that could arise at the hex plate. The flexure material properties were varied to determine the best trade of optical path stability, launch survivability, manufacturability, and cost. The materials considered were Aluminum (Al 6061-T6) and Titanium (Ti-6Al-4v).  For the structural analysis the hex plate was fixed at three points on its outer edge simulating spacecraft support, and MAC loads (100g accelerations in three orthogonal directions) were applied to the model.  Various flexure throat cross sectional areas were considered, and the best design was chosen based on the resulting maximum stresses revealed in the FEA.  The results for each material selection under this loading are shown in table \ref{tab:macload}. 

\begin{table}[H]
    \centering
    \begin{tabular}{c c c}
    \hline
     Load Case Direction & Max Flexure Stress [Al] & Max Flexure Stress [Ti] \\
    \hline
        100g $\hat{x}$ & 103ksi & 112ksi \\
        100g $\hat{y}$ & 98ksi & 103ksi \\ 
        100g $\hat{z}$ & 42ksi & 46ksi \\
    \hline
    \end{tabular}
    \caption{MAC loading in three orthogonal directions and the corresponding maximum flexure stress for Aluminum and Titanium flexures. The Ti flexures were chosen due to their higher maximum stress.}
    \label{tab:macload}
\end{table}

An additional analytical consideration for the structure is the modal behavior of the OTA.  The modal FEA reveals the lowest expected resonant frequency of the OTA under the prescribed mounting conditions for given material selections and flexure design choices.  The table below shows the first three modal FEA results for the same conditions discussed above.  These results indicate there would be little chance of exciting resonance in the OTA structure when considering expected launch loads.  And similarly, the telescope itself would not likely impart damaging resonance to any other spacecraft systems or partner satellites if launched as a secondary payload ride-share.

\begin{table}[H]
    \centering
    \begin{tabular}{c c c}
    \hline
        Al Flexures & Ti Flexures  \\
    \hline
        130Hz & 156Hz \\
        130Hz & 156Hz \\
        363Hz & 434Hz \\
    \hline
    \end{tabular}
    \caption{Modal behavior of the OTA for Al and Ti flexures for the cases described in \ref{tab:macload}. Based on the results from the FEA for both materials, the Titanium flexure design was selected based on the extra margin provided on mass and stress.}
    \label{tab:my_label}
\end{table}

 Verification  of the selected flexure design needed to also consider the thermal behavior of the system.  For ease of creating the thermal FEM, the boundary conditions for the structure were preserved from the structural FEA and a 27.8°C differential in temperature was applied to the hex plate.  This set of conditions would bring into focus the thermal isolation of the OTA from the hex plate.  The results from this thermal analysis were normalized to show an expected mirror sag of 10nm P-V for each 1.0°C of temperature differential.  The results (shown below) also indicate an axisymmetric deformation of the OTA as would be expected.

\begin{figure}[H]
    \centering
    \includegraphics[width=0.55\textwidth]{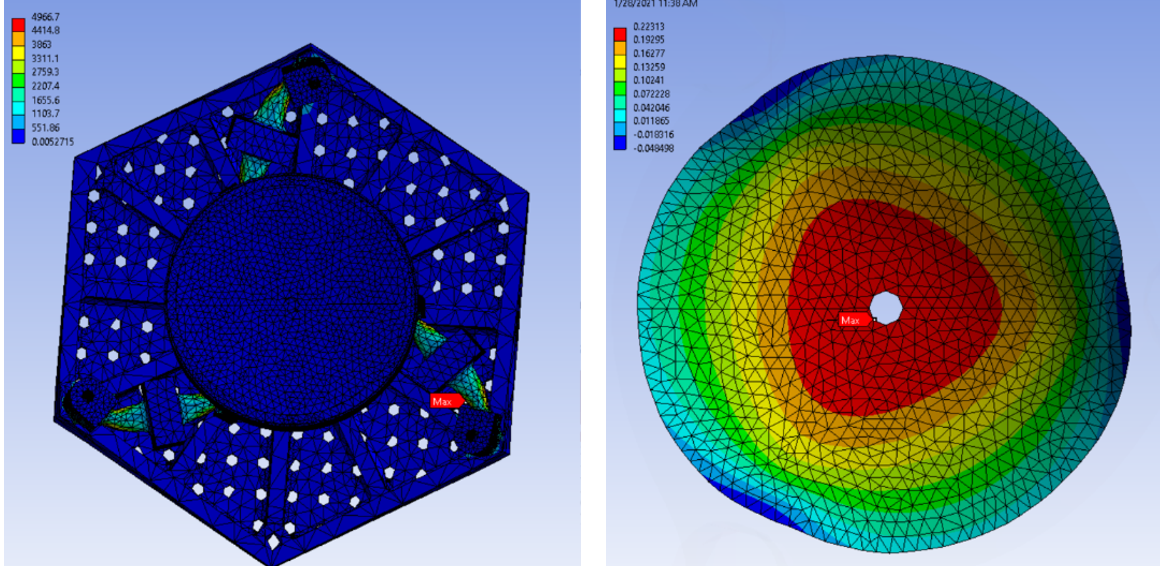}
    \caption{Finite Element results of the hexplate-primary mirror temperature differential. The flexure design results in $\approx$ 10nm of trefoil surface sag per degree Celsius.}
    \label{fig:fea_results}
\end{figure}

Overall, an iterative analysis process was implemented to determine the final telescope structure geometry.  The results from dynamic survivability, modal, and thermal analyses were considered for each element.  Part of the FEM definition involves application of material properties that also allow for the consideration of mass as a secondary factor in selection of the final geometry.  Resulting optical performance effects are then analyzed in the Zemax OpticStudio STAR module for each iteration by exporting the three-dimensional nodal parameters from the FEA results. Other geometry selections made during the design process were more straightforward than the flexural interface between the OTA and the hex plate.  For instance, given the axisymmetric behavior for all analyses, the tube geometry selections were made without direct optical analyses as the major performance effect would be in focus change due to varying thermal conditions.  For this, common engineering knowledge can be applied to bias the focus alignment for all optical elements prior to integration with an instrument for the given expected on-orbit operational conditions (for this project +/-1°C).  Similarly, the hex plate design only needed to preserve the axisymmetric geometry approximation when considering its overall shape and interface definition.  
While not discussed in detail earlier, mass of each components was considered as an important mechanical design decision.  An overall maximum allowable mass budget of 2 kg was allotted to the design and did not present a problem during the iterative analyses.  The final telescope design mass is expected to be less than 1 kg once all hardware is integrated.  The only major design decision based on mass reduction involved light-weighting the primary mirror blank by cutting pockets into the structure and analyzing these pockets for their structural effect on mirror surface figure.  

\begin{figure}[H]
    \centering
    \includegraphics[width=0.6\textwidth]{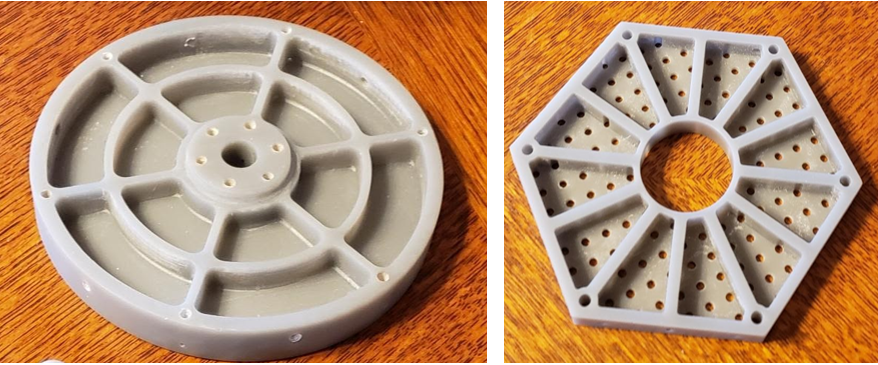}
    \caption{3D Prints of the primary mirror (left) and hex plate (right) demonstrating the lightweighting scheme used to achieve the mass requirement of $< 2kg$}
    \label{fig:my_label}
\end{figure}

\section{STOP EVALUATION - MERGING FEA WITH SEQUENTIAL RAYTRACING}

Numerous loading conditions were studied through FEA software to ensure the overall survival of the VCT payload, however, FEA analysis by itself could not determine how the deformed elements would impact the system’s optical performance. Further studies of the wavefront deformation were necessary to confirm that the system would be able to function properly in its environment of operation. Typically, this kind of analysis would require the deformed surface to be decomposed into hundreds of polynomial terms to accurately model the shape of the surface, with no guarantee of the raytracing software being able to support the number of terms needed. However, through use of OpticStudio’s STOP (Structural, Thermal, and Optical Performance) analysis feature \emph{STAR} (Structural, Thermal, Analysis and Results) evaluation of the system’s performance in various loading scenarios was made quick and efficient. STAR allows for the data from FEA software to be loaded directly into OpticStudio and onto select optical elements. By using the position and displacement of each node in a 6 column (x, y, z, dx, dy, dz) format STAR accurately deforms the surface shape, allowing for the deformed wavefront to be observed under various loading conditions. For the purposes of our design study we primarily looked at the surface deformation on the primary mirror caused by rotation about the optical axis.

\begin{figure}[H]
    \centering
    \includegraphics[width=\textwidth]{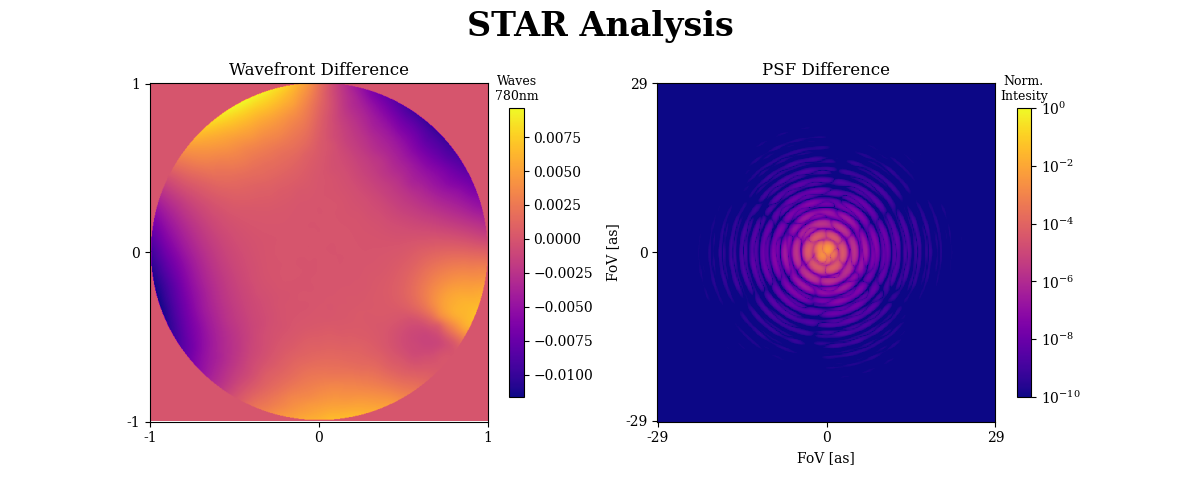}
    \caption{Influence of STAR on the wavefront (left) and the PSF (right). A force tangent to the primary mirror was applied in FEA and the resultant surface deformation was loaded into the OpticStudio raytrace model. This permits evaluation of performance degradation as a function of structural and thermal deformation.}
    \label{fig:star}
\end{figure}

In figure \ref{fig:star} the change of the optical system's wavefront and PSF are shown after applying a -237 arcsecond rotation about the optical axis. This causes clear degradation in the system's overall optical performance. This optical analysis is critical for an iterative design process between optical and mechanical design teams. Quick optical performance analysis through STAR allows for the mechanical team to adjust the structures supporting the optical surfaces where necessary, ensuring performance specifications are met in realistic structural and thermal scenarios. 

\section{Laboratory Prototype Status}

\begin{figure}[H]
    \centering
    \includegraphics[width=0.7\textwidth]{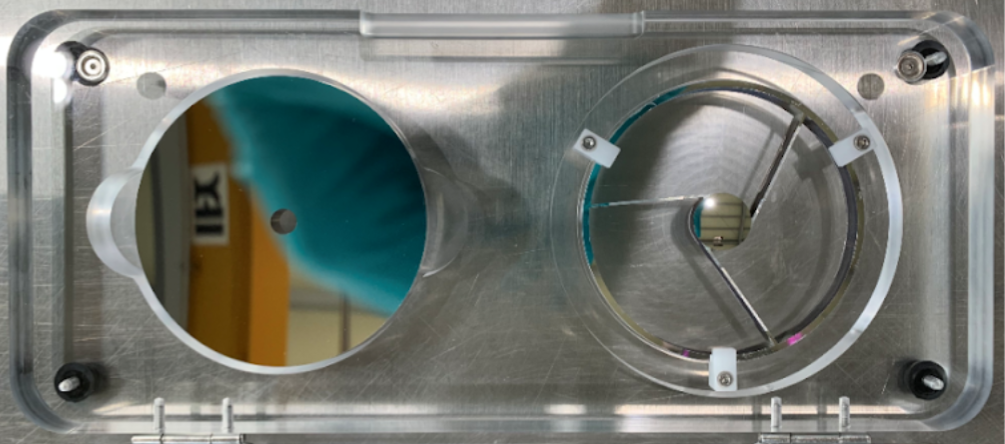}
    \caption{The fabricated (left) primary and (right) secondary of the on-axis VCT design manufactured by Hanbat National University in South Korea.}
    \label{fig:fab_mirrors}
\end{figure}

The next step for the development of the VCT is the construction of a laboratory prototype of the on-axis aluminum VCT. Mirrors for the on-axis VCT have been fabricated by Hanbat national university and are shown in figure \ref{fig:fab_mirrors}. 

The primary and secondary mirror have arrived at the University of Arizona and are awaiting measurement and assembly so that they can be mounted into the OTA described in section 5 for system-level tests. The UArizona Space Astrophysics lab test facilities include a TVAC chamber and Hexapod that can be used for thermal vaccum and jitter tests respectively. Details of the comprehensive characterization and assembly of the Versatile Cubesat Telescope prototype will be published in the future.

\section{Conclusion}

We present diffraction-limited on- and off-axis designs for a Versatile CubeSat Telescope that fits within a 2U volume at low cost. Comprehensive analysis of the sensitivity to misalignment and polarization were considered, and a stable mechanical housing was created for the on-axis VCT. The VCT is a well-characterized high-performance design that can be adapted to a variety of space-borne research payloads. Future iterations of the VCT will expand upon the development of MEMS devices in space (e.g. DeMi\cite{Morgan21}) by replacing the FSM with a Deformable Mirror for a higher degree of active wavefront control. Closed loop thermal simulations will be conducted to demonstrate the behavior of the VCT in response to a dynamic thermal environment.

\section{Acknowledgements}
We thank Zemax for the early access to their STAR analysis feature. This research made use of community-developed core Python packages, including: Numpy\cite{numpy}, Matplotlib \cite{matplotlib}, SciPy \cite{jones_scipy_2001}, Astropy \cite{the_astropy_collaboration_astropy_2013}, and Jupyter, IPython Interactive Computing architecture \cite{perez_ipython_2007,kluyver_jupyter_2016}. Portions of this work were supported by the Arizona Board of Regents Technology Research Initiative Fund (TRIF).

\bibliography{report} 
\bibliographystyle{spiebib} 

\end{document}